\documentclass[final]{aipproc}

\usepackage{natbib}
\usepackage{amsmath,amsfonts,amssymb,amsxtra}

\layoutstyle{6x9}

\begin{document}

\title{Modeling CHANDRA Low Energy Transmission Grating Spectrometer
       Observations of Classical Novae with \tt{PHOENIX}}


\author{A. Petz}{
  address={Hamburger Sternwarte, Gojenbergsweg 112, 21029 Hamburg,
           Germany; [apetz,phauschildt]@hs.uni-hamburg.de}}
\author{P.~H. Hauschildt}{
  address={Hamburger Sternwarte, Gojenbergsweg 112, 21029 Hamburg,
           Germany; [apetz,phauschildt]@hs.uni-hamburg.de}}
\author{J.~U. Ness}{
  address={Oxford University, Theoretical Physics, 1 Keble Road, Oxford,
           OX1 3NP, UK; ness@thphys.ox.ac.uk}}
\author{S. Starrfield}{
  address={Dept of Physics and Astronomy, ASU, P. O. Box 871504, Tempe,
           AZ 85287-1504; sumner.starrfield@asu.edu}}

\begin{abstract}
We use the {\tt PHOENIX} code package to model the X--ray spectrum of
Nova V4743 Sagittarii observed with the LETGS onboard the Chandra
satellite on 19 March 2003. To analyze nova atmospheres and related
systems with an underlying nuclear burning envelope at X--ray
wavelengths, it was necessary to update the code with new microphysics.
We demonstrate that the X--ray emission is dominated by thermal
bremsstrahlung and that the hard X--rays are dominated by Fe and N
absorption. Preliminary models are calculated assuming solar
abundances. It is shown that the models can be used to determine
element abundances in the nova ejecta by increasing the absorption in
the shell.
\end{abstract}

\maketitle

\section{Introduction}

We have modeled X--ray spectra of classical novae (CNe) with the {\tt
PHOENIX\rm}--code version 13 \cite[]{petz04}. With our models it is
possible to determine the effective temperature of the nova atmosphere
and the formation of X--ray emission in the nova shell. We show that it
is possible to determine the abundances of the ejecta by modelling
X--ray spectra of CNe and present our first results for the abundances
of C, N, and O.\par
For the comparison of our model with observations we use an X--ray
spectrum of Nova V4743 Sagittarii (Sgr) observed with the LETGS onboard
the Chandra satellite in March 2003 \cite[]{ness03}.

\section{Modeling nova atmospheres in X--rays with {\tt PHOENIX}}

Our model atmospheres are 1D spherical symetric, expanding, and time
independent. The radiative equilibrium is solved in the comoving frame
and the radiative transfer is treated as special relativistic for an
atmosphere in full NLTE with line blanketing of 8534 atomic levels from
H, He, C, N, O, Ne, Mg, and Fe. We use an exponential density law
($\rho(r) \propto r^{-n}$) with a gradient of $n = 3$ and a standard
velocity field ($v(r) = \dot{M} / 4 \pi r^2 \rho(r)$\,, $\dot{M}$ =
const) with an outer velocity of $v_{\rm out} = 2500$\,km s$^{-1}$.\par
The earlier versions of {\tt PHOENIX} used atomic data from CHIANTI
Version 3 (CHIANTI3) and the line lists of Kurucz. For this work, we
have implemented two new atomic databases, CHIANTI Version 4 (CHIANTI4)
\cite[]{young03} and
APED\footnote{\tt http://cxc.harvard.edu/atomdb/\rm}, because the
old databases did not provide enough data for the X--ray energy range.
Using the new databases, we have extended {\tt PHOENIX} to use many new
spectral lines in the X--ray waveband down to 1 {\AA}, improved data
for electron collision rates, new data for proton collision rates, and
better data for thermal bremsstrahlung.

\section{X--ray observations of nova V4743 Sgr}

The solid curve of Fig. 1 shows the observed X--ray spectrum of nova
V4743 Sgr on March 19, 2003 with the LETGS onboard the CHANDRA satellite
\cite[]{ness03}. At wavelengths greater than $\approx 55$\,{\AA}, it is
dominated by second and higher dispersion orders, and these wavelengths
will not be considered in our analysis. The effective areas used to
convert from ct s$^{-1}$ to flux are determined with the CIAO software
package\footnote{\tt http://cxc.harvard.edu/ciao/\rm}, version 3.0.\par
An examination of the spectrum shows that it is not a black--body but
resembles a stellar atmosphere with deep absorption features and,
possibly, some weak emission lines. The strongest lines are from the two
highest ionisation stages of C, N, and O. An extensive analysis of the
observation has been carried out by \cite{ness03}.

\section{Model with solar abundances}

The best fit with solar abundances to the spectrum of nova V4743 Sgr is
shown in Fig. 1, left panel. The model has an effective temperature of
$T_{\rm eff} = 5.8 \times 10^5$\,K and a bolometric luminosity of
$L_{\rm bol} = 50,000 L_\odot$. To get the correct slope for the
pseudo--continuum, a value of $n_h = 4.0 \times 10^{21}$\,cm$^{-2}$ for
the hydrogen column density has to be used. The quality of the fit is
independent of the luminosity of the model. This was already found for
earlier solar models in other wavelength ranges
\cite[]{hauschildt95b}.\par
Close inspection of the measured spectrum reveals that some spectral
lines are not reproduced well or are missing in the model spectrum. This
is because we have used only solar abundances in this model and have not
increased abundances of, for example the CNO elements, as is generally
observed in novae and predicted by theory
\cite[]{starrfield98}. Furthermore all absorption lines are too weak and
there is too much emission around $\lambda \sim 24$\,{\AA}. Increasing
the abundances should increase the absorption and the fit should improve
with solar abundances.\par
The X--ray emission is dominated by thermal bremsstrahlung from the
atmosphere surrounding the WD and the hard spectral range of
$\lambda \lesssim 29$\,{\AA} is dominated by iron and nitrogen
absorption. The atmosphere is in strong departure from LTE and is very
extended with the highest ionization stages of elements in the outest
layers.

\section{Model with non--solar abundances}

A model with non--solar abundances (Fig. 1, right panel) produces a
spectrum which fits the observation much better than with solar
abundances. It was calculated with a 22 times solar abundance of
nitrogen and oxygen. Accordingly the nitrogen to oxygen ratio is equal
to the solar value. The abundance of carbon with 1.25 times solar is
only slightly higher than in solar material.\par
The strengths of some spectral lines now fit much better to the
observation. For example, the fits of the N VII line at
$\lambda \sim 24.8$\,\AA{} and the O VII line at
$\lambda \sim 21.6$\,\AA{}. Around $\lambda \sim 24$\,{\AA} there is
still too much emission.

\begin{figure}
  \includegraphics[width=0.4\textwidth,angle=90]{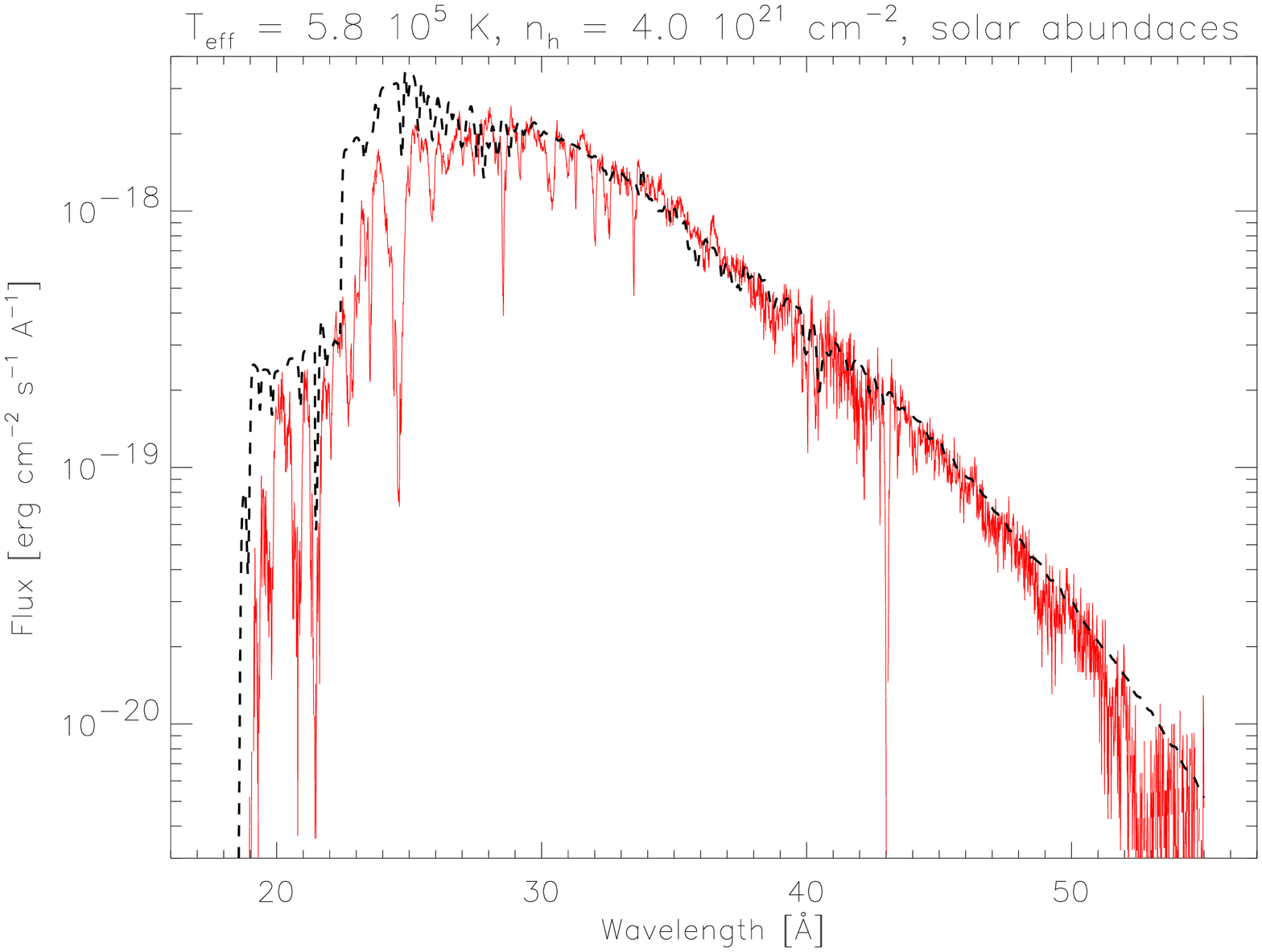}
  \includegraphics[width=0.4\textwidth,angle=90]{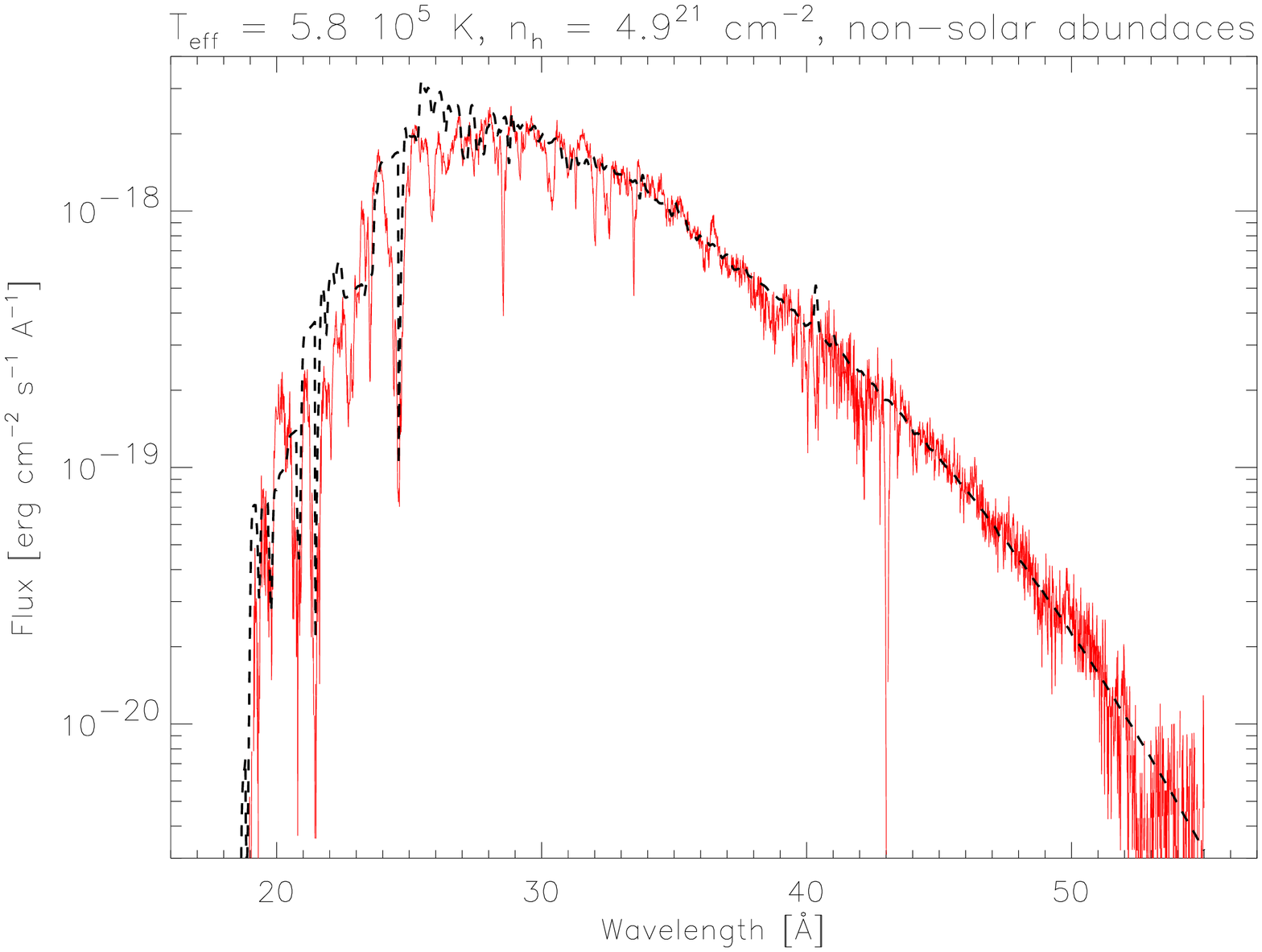}
  \caption{Model atmosphere of nova V4743 Sgr with solar (left) and
           non--solar abundances (right). The solid curve is the
           observed spectrum.}
\end{figure}

\begin{theacknowledgments}
Some of the calculations presented here were performed at the
H\"ochstleistungs Rechenzentrum Nord (HLRN) and at the National Energy
Research Supercomputer Center (NERSC), supported by the U.S. DOE. We
thank all these institutions for a generous allocation of computer
time. Part of this work was supported by the DFG
(\it Deutsche Forschungsgemeinschaft\rm), project number
HA 3457/2--1. S. Starrfield was partially supported by grants from
NASA--CHANDRA, NASA--Theory, and NSF to ASU.
\end{theacknowledgments}

\end{document}